\documentclass[letterpaper, 10pt, conference]{ieeeconf}
\IEEEoverridecommandlockouts
\usepackage{amsmath} 
\usepackage{amssymb}  
\usepackage{graphicx}
\usepackage{amsfonts}
\usepackage{float}
\usepackage{color}
\usepackage{xcolor}
\usepackage{subcaption}
\usepackage[font=footnotesize]{caption}
\usepackage{array}
\usepackage[markup=default]{changes}
\usepackage{booktabs,array,dcolumn}
\newcolumntype{P}[1]{>{\centering\arraybackslash}p{#1}}
\usepackage{mathtools}
\usepackage{algpseudocode}
\usepackage{listings}
\usepackage{cite}
\usepackage{color}
\usepackage{comment}
\usepackage[ruled,vlined]{algorithm2e}
\usepackage{graphicx}
\usepackage{color}
\usepackage{adjustbox}
\usepackage{centernot}
\usepackage{csquotes}
\usepackage{mathtools}
\usepackage{mathabx}
\usepackage{physics}
\usepackage{float}
\usepackage{multirow}
\usepackage{enumerate}
\usepackage{color}
\usepackage{cite}

\usepackage{xcolor}
\usepackage[colorlinks,citecolor=cyan,urlcolor=blue]{hyperref}
\usepackage{hyperref}
\usepackage{etoolbox}
\newcommand{\rom}[1]{\uppercase\expandafter{\romannumeral #1\relax}}

\usepackage{bm}
\usepackage{balance}
\usepackage{mathtools}

\newtheorem{theorem}{Theorem}
\newtheorem{corollary}{Corollary}
\newtheorem{remark}{Remark}

\newtheorem{result}{Result}

\usepackage{color}
\definecolor{gray}{RGB}{128,128,128}

\newcommand{\remove}[1]{}

\begin{document}
\title{Information-Theoretic Threshold Design for Sparse Event Encoding in Learning Systems}

\author{Pratishtha Shukla$^{1,4}$, Maksudul Alam$^{2,5}$, Srikanth B. Yoginath$^{2,6}$, and Ali R. Ekti$^{3,7}$\\
\small{$^{1}$Computational Sciences and Engineering Division, Oak Ridge National Laboratory,}\\
\small{$^{2}$Computer Science and Mathematics Division, Oak Ridge National Laboratory,}\\
\small{$^{3}$Electrification and Energy Infrastructures Division, Oak Ridge National Laboratory}\\
\small{Email: $^{4}$shuklap@ornl.gov, $^{5}$alamm@ornl.gov, $^{6}$yoginathsb@ornl.gov, $^{7}$ektia@ornl.gov}
\thanks{*This manuscript has been authored by UT-Battelle, LLC, under contract DE-AC05-00OR22725 with the U.S. Department of Energy (DOE). The U.S. government retains and the publisher, by accepting the article for publication, acknowledges that the U.S. government retains a nonexclusive, paid-up, irrevocable, worldwide license to publish or reproduce the published form of this manuscript, or allow others to do so, for U.S. government purposes. DOE will provide public access to these results of federally sponsored research in accordance with the DOE Public Access Plan (https://www.energy.gov/doe-public-access-plan).}
}
\maketitle

\begin{abstract}
One approach to event-driven learning is to encode continuous signals as sparse events generated when signal changes exceed a prescribed threshold. However, this threshold is typically selected empirically, with limited theoretical guidance. This paper establishes a connection between threshold-based event encoding and discretized learning systems and identifies a dimensionless ratio governing information retention. From this connection, we derive an information-sufficiency boundary using normal-condition data and formulate a data-driven criterion that selects the sparsest threshold retaining a prescribed fraction of peak discriminative information. The framework is illustrated using event cameras and evaluated on power-system fault classification. Results show that the proposed design identifies sparse operating thresholds while exposing the tradeoff between information retention, event activity, and classification performance.
\end{abstract}

\section{Introduction}
\label{sec:intro}
Event-driven sensing and neural computation have emerged as a promising paradigm for low-power, low-latency edge intelligence. Spiking neural networks (SNNs) and neuromorphic architectures process information as discrete spike events rather than dense continuous signals, enabling sparse computation and reduced communication overhead~\cite{nilsson2023integration, yao2024spike, m2025comparative}. A common component of event-driven systems is an \emph{event encoder}, which converts a continuous measurement stream into a sparse event stream by generating events when signal changes exceed a prescribed threshold. This delta-thresholding mechanism appears across event cameras~\cite{gallego2020event, cimarelli2025hardware}, neuromorphic sensing and edge-computing systems~\cite{yao2024spike, nilsson2023integration}, and spike-based classifiers for streaming physical signals~\cite{tian2023neurocare}.

Despite its central role, the encoding threshold is commonly selected through empirical search or manual tuning, often requiring repeated evaluation of the learning model over candidate thresholds and providing limited guidance that transfers across models or datasets. This choice introduces a tradeoff between event sparsity and information retention. A threshold that is too low can generate excessive events from normal signal variation, increasing communication and event-processing demands, whereas a threshold that is too high can suppress changes that are informative for subsequent learning.

This paper addresses this gap by establishing a formal isomorphism between threshold-based event encoding and discretization in input-output learning systems, a setting for which information-theoretic learnability results are available \cite{shukla2026finite}. Under this correspondence, we identify an information-sufficiency boundary from normal-condition signal statistics. Rather than treating this boundary as the operating threshold, it defines the lower boundary of the threshold-design region. 

Building on this result, we formulate threshold selection as a tradeoff between information retention and event sparsity. Candidate thresholds above the boundary are evaluated according to the discriminative information retained in the event representation, and the operating threshold $\theta^*$ is selected as the largest threshold that retains a specified fraction of the peak attainable information. This provides a principled way to obtain a sparse event representation without relying solely on empirical threshold tuning.

The proposed approach is first illustrated using event cameras, where delta-thresholding is inherent to the sensing mechanism. We then evaluate the threshold-design method using power-system transient measurements, where continuous voltage and current waveforms are delta-threshold encoded into temporal event trains for a Spiking Neural Network (SNN)-based Convolutional Variational Autoencoder (CVAE) fault classifier~\cite{alam2024convolutional, khodayar2023deep}. Such event-driven architectures offer the potential for sparse, low-latency, and energy-efficient edge processing~\cite{pfeiffer2018deep}.

\section{Threshold Design Theory}
\label{sec:theory}

Threshold-based event encoding determines what information from a continuous signal is made available to a learning system. Changes below the encoding threshold are discarded, while changes above it are retained as discrete events. The choice of threshold therefore controls both the sparsity and the information content of the resulting representation: a small threshold retains more signal variation but may also encode normal-condition fluctuations, whereas a large threshold suppresses this activity but may discard changes relevant to the learning task.

This tradeoff can be analyzed by interpreting the encoding threshold as a resolution scale. From this perspective, event encoding is closely related to discretization, where continuous-valued observations are represented at a finite resolution. We first recall an information-limit result for discretized learning systems and show that signed delta encoding has the same mathematical structure. This correspondence yields a dimensionless diagnostic for event encoding and identifies a lower boundary for information-sufficient operation. We then use this boundary to formulate threshold selection as a tradeoff between event sparsity and retained discriminative information.

\subsection{Information Limit Under Discretization}
\label{sec:fir}

We begin with the discretized learning setting of \cite{shukla2026finite}, which provides the theoretical basis for
the event-encoding analysis developed below.

Consider a scalar input-output system in which the output $Y_t$ is discretized into $q$ uniform bins of width $\Delta=1/q$, while measurement noise $\varepsilon_t\sim\mathcal{N}(0,\sigma_\varepsilon^2)$ perturbs the output before discretization. A predictive model selects input variables that best explain the discretized output through the coarsened conditional entropy $H_q(M)$.

The key quantity is not the noise magnitude or discretization resolution individually, but their ratio
\begin{equation}
    \rho =
    \frac{\sigma_\varepsilon}{\Delta}
    = \sigma_\varepsilon q,
    \label{eq:rho_fir}
\end{equation}
which measures stochastic variability relative to the resolution at which the system is represented.

\begin{result} \label{result:fir}
    The following conditions summarize the phase transition of information regimes \cite{shukla2026finite}.
\begin{enumerate}[(i)]
    \item \textbf{Resolved regime} $(\rho < 1)$: conditional output distributions remain distinguishable
    across cells and entropy-based learning is asymptotically consistent.

    \item \textbf{Saturation regime} $(\rho > 1)$: stochastic variability exceeds the discretization scale,
    causing the conditional distributions to lose distinguishability and entropy-based discrimination to fail.

    \item \textbf{Boundary regime} $(\rho \approx 1)$: the system transitions between these two behaviors.
\end{enumerate}
\end{result}

The key feature of this result is that the transition between the two regimes is determined by the ratio of stochastic variability \(\sigma_\epsilon\) to the discretization resolution \(\Delta\). Threshold-based event encoding has an analogous structure: normal signal variability determines the scale of fluctuations presented to the encoder, while the encoding threshold determines the resolution at which those changes are represented. We next formalize this correspondence.

\subsection{Event Encoding as Discretization}
\label{sec:isomorphism}

Let $x(t)$ denote a scalar waveform channel sampled at $t=1,\ldots,T$, and define its temporal gradient as
\begin{equation}
    \Delta x(t)=x(t)-x(t-1).
\end{equation}
The variability of \(\Delta x(t)\) under normal operating conditions provides the reference scale for background signal fluctuations, while changes exceeding this scale may contain event-related information. A signed delta encoder represents these changes through positive and negative event channels,
\begin{equation}
    s^+(t)=\mathbf{1}\{\Delta x(t)>+\theta\},\quad
    s^-(t)=\mathbf{1}\{\Delta x(t)<-\theta\},
    \label{eq:encoder}
\end{equation}
where $\theta>0$ is the encoding threshold and \(\mathbf{1}\{\cdot\}\) is the indicator function, equal to 1 when the enclosed condition holds and 0 otherwise.

The threshold $\theta$ determines which changes are retained: gradients with $|\Delta x(t)|\leq\theta$ produce no event, whereas changes exceeding the threshold are represented by positive or negative events. The relevant question is therefore how $\theta$ compares with the natural variability of $\Delta x(t)$ under normal conditions.

\begin{theorem}[Event-Encoding Isomorphism]
\label{thm:isomorphism}
The signed delta encoder in~\eqref{eq:encoder} is formally
isomorphic to a symmetric three-region discretization of
$\Delta x(t)$, where the encoding threshold $\theta$ corresponds to
the discretization resolution $\Delta$. Defining the standard deviation of normal-condition signal increments as \(\sigma_\Delta\), the corresponding dimensionless variability-to-threshold ratio is
\begin{equation}
    \rho_E=\frac{\sigma_\Delta}{\theta},
    \qquad
    \sigma_\Delta =
    \mathrm{std}(\Delta x\mid\mathrm{normal}).
\end{equation}
This has the same variability-to-resolution form as
$\rho=\sigma_\varepsilon/\Delta$ in the discretized learning system \cite{shukla2026finite}. No distributional assumption on $\Delta x$ is required
for this correspondence.
\end{theorem}

\begin{proof}
A symmetric three-region discretization partitions a scalar variable
$z$ as
\[
B_-:z<-\Delta,\qquad
B_0:|z|\leq\Delta,\qquad
B_+:z>\Delta.
\]
Setting $z=\Delta x(t)$ and $\Delta=\theta$ gives
$B_-\Leftrightarrow s^-(t)=1$,
$B_0\Leftrightarrow s^+(t)=s^-(t)=0$, and
$B_+\Leftrightarrow s^+(t)=1$.
The two representations therefore induce identical partitions.
\end{proof}
The significance of Theorem~\ref{thm:isomorphism} is that the encoding threshold can be
interpreted relative to the natural variability of the signal rather
than as an arbitrary hyperparameter. Under the interpretation of
$\sigma_\Delta$ as the stochastic variability presented to the
encoder, $\rho_E=\sigma_\Delta/\theta$ plays the same role as the
variability-to-resolution ratio in Result~1. This correspondence
motivates the following event-encoding regimes.


\begin{corollary}[Event-Encoding Information Regimes]
\label{cor:regimes}
For the signed delta encoder:
\begin{enumerate}[(i)]
    \item $\rho_E < 1$: the encoder operates in the information-sufficient regime, where normal-condition fluctuations are largely suppressed relative to the encoding threshold;

    \item $\rho_E > 1$: the encoder operates in the information-limited regime, where normal-condition fluctuations increasingly cross the threshold and can obscure task-relevant distinctions.
\end{enumerate}
\end{corollary}

This limitation occurs before any learning model or classifier is applied. For the processing chain $x \rightarrow \Delta x \rightarrow \mathbf{s}
\rightarrow f(\mathbf{s})$, the data-processing inequality gives
\begin{equation}
    I\bigl(Y;f(\mathbf{s})\bigr)
    \leq I(Y;\mathbf{s})
    \leq I(Y;\Delta x)
    \leq I(Y;x),
    \label{eq:info_chain}
\end{equation}
where $Y$ denotes the class label and $I(A;B)$ is the mutual information between random variables $A$ and $B$, defined as
\begin{equation}
    I(A;B)=
    \sum_{a,b}p(a,b)
    \log\frac{p(a,b)}{p(a)p(b)}.
    \label{eq:mutual_information}
\end{equation}
Information removed by the encoder cannot be recovered by increasing the capacity of the learning model. A classifier may exploit redundancy across channels or time, but cannot recover class information absent from the encoded representation. Thus, \(\rho_E\) characterizes the information quality of the representation rather than the accuracy of a particular classifier. The excess-risk result of \cite{shukla2026finite} can also be expressed under this correspondence, subject to the assumptions discussed below.

\noindent\textbf{Transferred Excess-Risk Bound:}
Under the event-encoding correspondence of Theorem~1, the excess-risk
bound of~\cite{shukla2026finite} takes the form
\begin{equation}
    \delta \geq
    \frac{\sigma_\Delta^2}{N_s}
    \left(K(M_{\mathrm{enc}})-K(M^*)\right),
\end{equation}
where $N_s$ is the number of training samples,
$K(M_{\mathrm{enc}})$ is the complexity measure associated
with the encoded representation, and $K(M^*)$ is the
corresponding measure for the reference model.

This expression is obtained by substituting $\sigma_\Delta$ for
$\sigma_\varepsilon$ under the correspondence in Theorem~\ref{thm:isomorphism}. The
original bound in~\cite{shukla2026finite} is derived for entropy-based
structure identification; its use here therefore provides a
transferred representation-level bound under the assumption that the
encoded event representation plays an analogous role to the
discretized output in that setting. For temporally correlated event
streams, the independence assumption underlying the equality form is
not satisfied, giving the stated lower-bound form.



Together, these results show that the encoding threshold determines the information available to the learning model before classifier training begins. This motivates identifying the threshold at which normal-condition variability becomes resolved by the encoder.

\subsection{Information-Sufficiency Boundary}
\label{sec:info_bound}
The boundary between the two encoding regimes occurs at $\rho_E = 1$. We therefore define the information-sufficiency threshold as
\vspace{-0.3cm}
\begin{equation}
    \theta_{\mathrm{IS}} = \sigma_\Delta.
    \label{eq:theta_is}
\end{equation}
Thresholds $\theta < \theta_{\mathrm{IS}}$ lie in the information-limited regime, while $\theta \geq \theta_{\mathrm{IS}}$ defines the admissible region for threshold design.

A useful feature of \eqref{eq:theta_is} is that $\theta_{\mathrm{IS}}$ depends only on the normal-condition gradient distribution. 
\begin{equation}
    \hat{\theta}_{\mathrm{IS}}
    =\mathrm{std}(\Delta x\mid\mathrm{normal}).
\end{equation}
The boundary can therefore be estimated directly from unlabeled data, without training a classifier.

The boundary, however, should not be interpreted as the recommended operating threshold. To see why, consider the event activity produced by normal-condition fluctuations at
and above this boundary. If these fluctuations are modeled as Gaussian, $\Delta x(t)\sim\mathcal{N}(0,\sigma_\Delta^2)$, the expected event rate is
\begin{align}
    C(\theta)
    &= \mathbb{P}(|\Delta x| > \theta) \notag \\
    &= \mathbb{P}(\Delta x > \theta)
     + \mathbb{P}(\Delta x < -\theta) \notag \\
    &= 2\left(1 - \Phi\!\left(\frac{\theta}{\sigma_\Delta}
       \right)\right),
    \label{eq:comm_cost}
\end{align}
where $\Phi$ is the standard normal CDF. At the information-sufficiency boundary,
\begin{equation}
    C(\theta_{\mathrm{IS}})
    =2\bigl(1-\Phi(1)\bigr)
    \approx0.317.
    \label{eq:rate_at_boundary}
\end{equation}

Thus, even at the onset of the information-sufficient regime, approximately $31.7\%$ of normal-condition samples would generate events under the Gaussian approximation. Although $\theta_{\mathrm{IS}}$ defines the theoretical boundary, it may therefore produce unnecessarily high event activity in practice. A practical operating point must additionally account for the tradeoff between event sparsity and retained task information.

\subsection{Information-Aware Threshold Design}
\label{sec:design}

This analysis motivates selecting a threshold that reduces event activity without discarding information needed for the subsequent learning task. These competing effects can be quantified directly. 
Event activity is measured by $C(\theta)$, while task-relevant information is measured by the mutual information (MI) between the class label $Y$ and the encoded representation $S_\theta$, as defined in~\eqref{eq:mutual_information}. Here,
$I(Y;S_\theta)$ quantifies the reduction in uncertainty about the class label provided by the encoded representation. Let the maximum discriminative information attainable over the admissible threshold range be
\begin{equation}
    I_{\max} = \max_{\theta\geq\theta_{\mathrm{IS}}} I(Y;S_\theta).
    \label{eq:I_max}
\end{equation}

Rather than maximizing information alone, we seek the sparsest representation that retains a prescribed fraction of this peak. The threshold-design problem is therefore
\begin{equation}
\begin{aligned}
    \min_{\theta}\quad & C(\theta)\\
    \mathrm{s.t.}\quad
        &\theta\geq\sigma_\Delta,\\
        &I(Y;S_\theta)\geq\eta I_{\max},
\end{aligned}
\label{eq:opt}
\end{equation}
where $\eta\in(0,1]$ is an application-dependent
information-retention requirement. It controls how aggressively the
encoder may trade information for sparsity. Larger values of $\eta$
require the selected threshold to remain closer to the
high-information region, whereas smaller values permit more aggressive
event suppression. Thus, $\eta$ provides a design parameter through
which the information requirement of a particular application can be
incorporated without tuning the threshold using classifier performance.

Since $C(\theta)$ decreases monotonically with $\theta$, the solution has a simple interpretation: $\theta^{\eta}$ is the \emph{largest} threshold that remains in the information-sufficient region while retaining at least $\eta I_{\max}$ of the peak discriminative information. Equivalently, it is the sparsest admissible event representation under the chosen information-retention
requirement.

Three threshold values therefore have distinct roles: \(\theta_{\mathrm{IS}}\) defines the information-sufficiency boundary, \(\theta_{I_{\max}}\) maximizes the retained class information, and \(\theta^{\eta}\) gives the sparsest representation satisfying the prescribed information-retention criterion. Under the design in \eqref{eq:opt}, these thresholds satisfy
\vspace{-0.3cm}
\begin{equation}
    \theta_{\mathrm{IS}}
    \leq \theta_{I_{\max}}
    \leq \theta^{\eta}.
\end{equation}

This ordering reflects the two stages of the threshold design. The information-sufficiency boundary defines the admissible region, while \(\theta_{I_{\max}}\) identifies the threshold of maximum discriminative information. Since the event rate \(C(\theta)\) decreases monotonically with \(\theta\), increasing the threshold beyond \(\theta_{I_{\max}}\) produces progressively sparser representations. The largest threshold that still retains at least \(\eta I_{\max}\) information is therefore selected as \(\theta^{\eta}\).


\subsubsection{Practical Threshold Computation}

For finite data, the optimization in \eqref{eq:opt} can be implemented by evaluating a discrete set of candidate thresholds. For thresholds \(\theta_1<\cdots<\theta_K\), with \(\theta_k\geq\theta_{\mathrm{IS}}\), the event representation \(S_{\theta_k}\) is generated at each threshold, and its class-discriminative information is estimated from the resulting event features as \(\hat I(Y;S_{\theta_k})\). The maximum over the candidate thresholds gives \(\hat I_{\max}\), and the operating threshold is selected as
\begin{equation}
    \hat{\theta}^{\eta} =
    \max\left\{
        \theta_k:
        \hat I(Y;S_{\theta_k})
        \geq\eta\hat I_{\max}
    \right\}.
    \label{eq:theta_hat}
\end{equation}

Thus, threshold selection reduces to a one-dimensional scan of the encoded data rather than repeated optimization of a learning model. The selected $\hat{\theta}^*$ is the sparsest tested encoding that retains the prescribed fraction of the peak estimated class information.

A closed-form solution would require an analytical characterization of $I(Y;S_\theta)$ under the class-conditional gradient distributions, which is left for future work.

\section{Illustrative Example: Event Camera Encoding}
\label{sec:event_camera}
To illustrate the physical interpretation of the proposed framework, we consider event cameras, where threshold-based event generation is a fundamental sensing mechanism. Event cameras provide a useful example because their sensing principle directly implements the delta-threshold encoding studied in Section~\ref{sec:theory}. This example demonstrates how the information-sufficiency condition maps to an established event-driven sensing system.

Event cameras respond to changes in scene intensity rather than recording complete image frames at fixed sampling intervals. At pixel $(u,v)$, an event is generated when the change in log intensity $\Delta L_{u,v}(t)$ exceeds a contrast threshold $C_{\mathrm{th}}$~\cite{gallego2020event,cimarelli2025hardware}:
\vspace{-0.2cm}
\begin{equation}
    e_{u,v}^{\pm}(t)
    =
    \mathbf{1}\{\pm\Delta L_{u,v}(t)>C_{\mathrm{th}}\}.
    \label{eq:event_camera}
\end{equation}
The event polarity indicates the direction of the intensity change. Thus, the sensor itself performs signed delta-threshold encoding before subsequent processing or learning.

Equation~\eqref{eq:event_camera} has the same structure as \eqref{eq:encoder}, with $\Delta x\equiv\Delta L$ and $\theta\equiv C_{\mathrm{th}}$. The corresponding dimensionless ratio is
\begin{equation}
    \rho_E =
    \frac{\sigma_{\Delta L}^{\mathrm{static}}}{C_{\mathrm{th}}},
    \label{eq:event_camera_rho}
\end{equation}
where $\sigma_{\Delta L}^{\mathrm{static}}$ characterizes measured variability under static-scene conditions. The information-sufficiency boundary therefore becomes
\begin{equation}
    C_{\mathrm{IS}}    =
    \sigma_{\Delta L}^{\mathrm{static}},
    \label{eq:event_camera_threshold}
\end{equation}
corresponding to $\rho_E=1$. Below this boundary, static-scene variability can generate background events~\cite{cimarelli2025hardware}. Increasing $C_{\mathrm{th}}$ suppresses such activity, while an excessively large threshold may also suppress task-relevant changes. The event-camera example therefore illustrates the information--sparsity tradeoff of Section~\ref{sec:design} in a physical event-driven sensing mechanism.

We next evaluate the framework quantitatively using power-system transient measurements in terms of event density, retained information, and classification performance.

\section{Power-System Application}
\label{sec:dataset}
Near real-time transient fault detection in three-phase power systems requires rapid identification of abnormal voltage and current behavior following events such as phase-to-phase and phase-to-ground short circuits. In a three-phase system, faults are commonly distinguished by the phases involved and by whether the fault includes a connection to ground. For example, AB, BC, and CA denote phase-to-phase faults; AG, BG, and CG denote single-phase-to-ground faults; ABG, BCG, and CAG denote two-phase-to-ground faults; and ABC and ABCG represent three-phase faults without and with ground involvement, respectively. These faults produce distinct transient signatures whose continuous processing can be computationally demanding for rapid detection and classification. Event-based processing instead represents signal changes as sparse event trains, making continuous-to-spike encoding a first step toward efficient fault detection and classification using SNN models.

The dataset used in this work is taken from \cite{alam2024convolutional} and contains these eleven fault classes together with normal operating conditions. Each class contains 80 simulated instances, for a total of 960 instances. Each instance consists of six time-series signals corresponding to the three-phase voltages \((V_A,V_B,V_C)\) and currents \((I_A,I_B,I_C)\). The instances span different fault types and operating conditions, providing variation both within and across classes.

For each physical channel $c$, the temporal change
\begin{equation}
    \Delta x_c(t)=x_c(t)-x_c(t-1)
\end{equation}
is computed after preprocessing and normalization. For the power-system experiments, signed delta encoding is used, with separate events generated for positive and negative signal changes
\begin{equation}
    S^{+}_{c,\theta}(t)=\mathbf{1}\{\Delta x_c(t)>\theta\},
    \quad
    S^{-}_{c,\theta}(t)=\mathbf{1}\{\Delta x_c(t)<-\theta\}.
    \label{eq:power_event_encoder}
\end{equation}
The six measured voltage and current channels therefore produce
12 binary event channels, preserving both the occurrence and direction of signal changes that exceed the encoding threshold. The following experiments first apply the proposed threshold-design procedure and then evaluate the resulting normalized operating points in the fault-classification task.

\subsection{Threshold Design Results}
\label{sec:mi_results}

The proposed design procedure selects the encoding threshold from the information retained by the event representation and its associated event activity, without training or evaluating the classifier during threshold selection. For this analysis, a window containing 4{,}000 pre-fault and 8{,}000 post-fault samples is extracted from each instance and downsampled by a factor of five. Each channel is standardized using statistics estimated from normal-condition data only, and no synthetic measurement noise is added. The temporal gradients are then computed from the standardized signals.

Normal-condition gradients provide the reference variability. For this experiment, $\sigma_\Delta=0.0942$, giving $\theta_{\mathrm{IS}}=0.0942$. Candidate thresholds are expressed as multiples of $\sigma_\Delta$. 

\begin{figure}[t]
    \centering
    \includegraphics[width=\columnwidth]{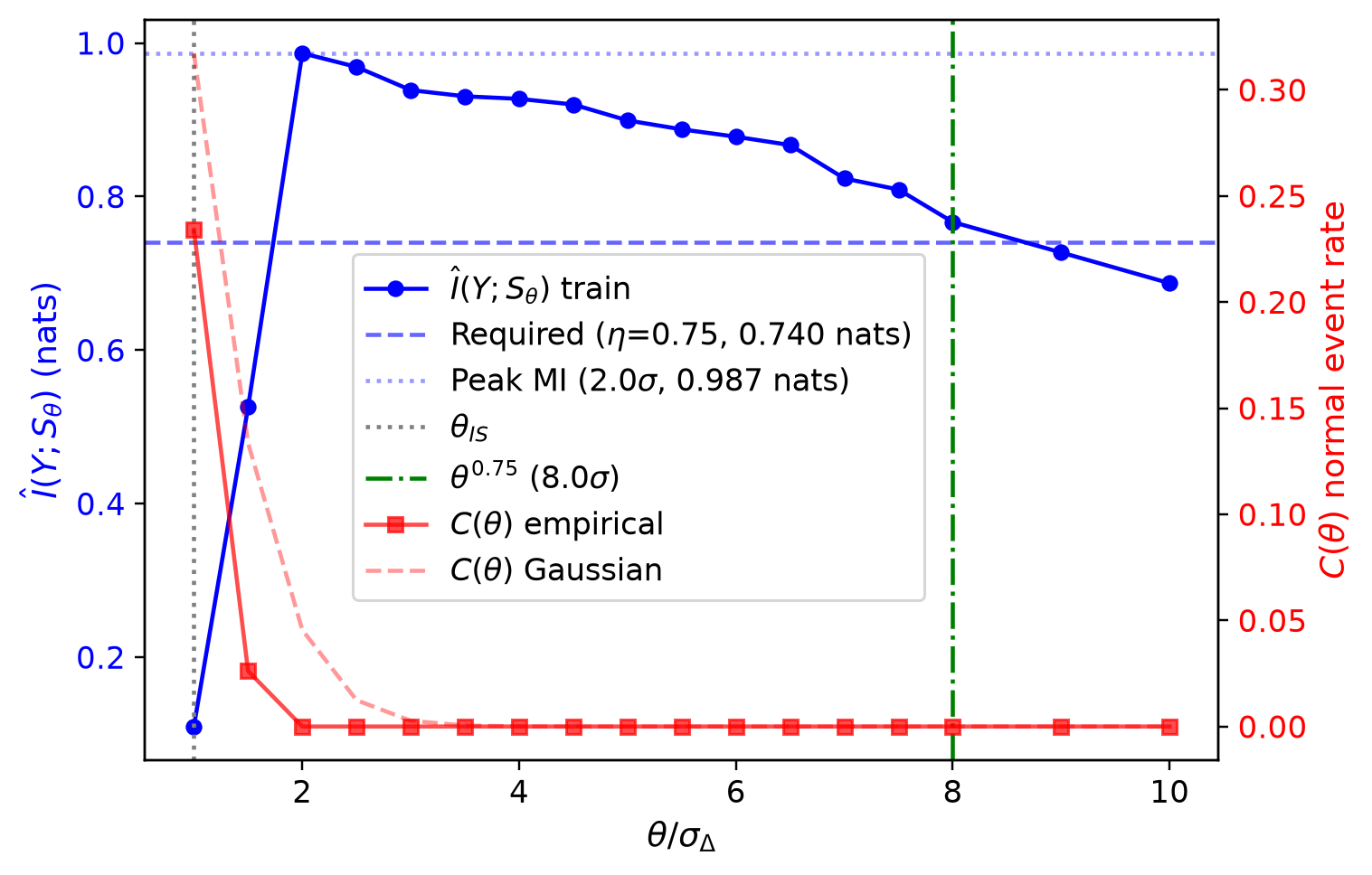}
    \caption{Information--sparsity tradeoff across normalized encoding thresholds.}
    \label{fig:mi_tradeoff}
\end{figure}

For each candidate threshold, the gradients are encoded using
\eqref{eq:power_event_encoder}. To obtain a compact discrete representation for empirical MI estimation, each instance is summarized by a 12-bit activity pattern, where each bit indicates whether the mean event rate of the corresponding channel exceeds 5\% over the analysis window. This pattern defines the discrete event representation \(S_\theta\). For this dataset, \(|\mathcal{Y}|=12\), corresponding to eleven fault classes and the normal operating condition, while \(|\mathcal{S}|=2^{12}\) represents the set of possible 12-bit activity patterns. MI \(\hat I(Y;S_\theta)\) is then estimated from the empirical joint and marginal frequencies of the class labels \(Y\) and event-pattern codes \(S_\theta\) over the training data. Repeating this calculation across candidate encoding thresholds gives the information curve in Fig.~\ref{fig:mi_tradeoff}.

\begin{remark}
The activity cutoff used to construct $S_\theta$ is a representation parameter and thus affects the estimated MI and the numerical value of $\theta^{\eta}$. As shown in Table~\ref{tab:activity_sensitivity}, across cutoffs from 0.25\% to 10\%, the peak-MI threshold remains within $2\sigma_\Delta$--$3\sigma_\Delta$, consistently above the theoretical information-sufficiency boundary, although the sparsity-selected $\theta^{\eta}$ varies with the representation. At larger cutoffs (15\%-20\%), the representation becomes increasingly restrictive, leaving only one or two candidate thresholds with sufficient activity for MI estimation. The threshold-selection procedure itself remains unchanged for any representation.
\end{remark}

\begin{table}[!t]
\centering
\caption{Sensitivity to the activity cutoff used to construct $S_\theta$.}
\label{tab:activity_sensitivity}
\footnotesize
\setlength{\tabcolsep}{3pt}
\begin{tabular}{|c|c|c|c|}
\hline
Activity cutoff &
$\theta_{I_{max}}/\sigma_\Delta$ &
$\theta^{\eta}/\sigma_\Delta$ &
MI at $\theta^{\eta}$ (nats) \\
\hline
0.25\% & 2.0 & 10.0 & 1.437 \\
0.5\%  & 2.5 & 10.0 & 1.421 \\
1.0\%  & 3.0 & 10.0 & 1.390 \\
2.5\%  & 2.5 & 10.0 & 0.991 \\
5.0\%  & 2.0 & 8.0  & 0.767 \\
7.5\%  & 2.0 & 6.0  & 0.530 \\
10.0\% & 2.0 & 4.5  & 0.302 \\
\hline
\end{tabular}
\end{table}

As shown in Fig.~\ref{fig:mi_tradeoff}, the estimated MI $\hat{I}(Y;S_\theta)$ initially increases with the threshold and reaches its maximum of 0.987~nats at $\theta=2\sigma_\Delta$. This behavior suggests that initially suppressing small signal variations removes event activity that contributes little to class discrimination. Beyond this point, $\hat{I}(Y;S_\theta)$ gradually decreases as increasing threshold values remove additional signal changes. At the same time, the empirical normal-condition event rate $C(\theta)$ decreases from 23.4\% at $\theta_{\mathrm{IS}}=\sigma_\Delta$ to below 1\% at $2\sigma_\Delta$ and below 0.01\% at $5\sigma_\Delta$ and $8\sigma_\Delta$.

To select an operating threshold from the tradeoff in Fig.~\ref{fig:mi_tradeoff}, we set the information-retention requirement to \(\eta=0.75\), meaning that the encoded representation must retain at least 75\% of the peak estimated class information. With \(\hat I_{\max}=0.987\) nats, the minimum required information is
$    \eta\hat{I}_{\max}
    =0.75(0.987)=0.740~\text{nats}.
$
Among the candidate thresholds satisfying this requirement, the largest admissible threshold is \(8\sigma_\Delta\). Because the design objective in \eqref{eq:opt} favors the sparsest representation satisfying the information constraint, the selected operating point is
\begin{equation}
    \theta^{0.75}=8\sigma_\Delta,
    \qquad
    \rho_E^{0.75}=\frac{\sigma_\Delta}{\theta^{0.75}}=0.125.
\end{equation}
At this threshold, $\hat I(Y;S_{\theta^{0.75}})=0.767$ nats,
corresponding to 77.7\% of the peak estimated mutual information.
For comparison, our previous trial-and-error threshold search required
repeated classifier evaluation and resulted in a threshold of
$5\sigma_\Delta$, within the information-retaining region identified
by the proposed method. Table~\ref{tab:mi_design} summarizes the
principal operating points identified by the design procedure.

\begin{table}[t]
\centering
\caption{Information-aware threshold design results. $\sigma_\Delta=0.0942$, $\eta=0.75$, and $\hat{I}_{\max}=0.987$~nats.}
\label{tab:mi_design}
\setlength{\tabcolsep}{3pt}
\small
\begin{tabular}{|l|c|c|c|c|}
\hline
Setting &
$\theta/\sigma_\Delta$ &
$\rho_E$ &
$\hat{I}$ (nats) &
Normal event
rate \\
\hline
$\theta_{\mathrm{IS}}$ &
1.0 & 1.000 & 0.110 & 0.234 \\
$\theta_{I_{max}}$ &
2.0 & 0.500 & 0.987 & $<10^{-2}$ \\
$\theta_{trial}$ &
5.0 & 0.200 & 0.899 & $<10^{-4}$ \\
$\theta^{\eta}$ &
8.0 & 0.125 & 0.767 & $<10^{-4}$ \\
\hline
\end{tabular}
\end{table}

Fig.~\ref{fig:mi_tradeoff} also shows the Gaussian null approximation \(C(\theta)=2[1-\Phi(\theta/\sigma_\Delta)]\). The empirical normal-condition event rate decreases more rapidly than this approximation, consistent with lighter tails in the observed normal-condition gradient distribution. Thus, for this dataset and the thresholds considered, the Gaussian approximation provides a conservative estimate of normal-condition event activity.

\subsection{Fault-Classification Performance}
\label{sec:cvae_results}

The operating points of Section~\ref{sec:mi_results} are evaluated on a downstream fault-classification task using an SNN-based CVAE architecture adapted from~\cite{alam2024convolutional}. In \cite{alam2024convolutional}, the CVAE was trained without labels on continuous voltage and current waveforms from the same dataset, with fault types forming separable clusters in the learned latent space.
The spiking architecture replaces continuous activations with leaky integrate-and-fire (LIF) neurons implemented using snnTorch~\cite{eshraghian2023snntorch}, and the continuous inputs are replaced by the temporal event trains from \eqref{eq:power_event_encoder}. For classification, the latent sampling layer and decoder are removed and a classification layer is attached to the encoder output. 

Because temporal sampling differs from that used for threshold design, $\sigma_\Delta$ is estimated independently for the classifier, giving $\sigma_\Delta^{\mathrm{clf}}=0.019324$. Since the encoding regime is determined by the normalized threshold $\theta/\sigma_\Delta$ rather than its absolute value, the operating points are transferred as $\theta/\sigma_\Delta\in\{1,2,5,8\}$, preserving their information-regime interpretation across the two sampling procedures. The classifier architecture and training are held fixed across experiments, with no classifier-based threshold retuning.

\begin{table}[!t]
\centering
\caption{Classification performance and event sparsity at the
nxormalized threshold operating points.}
\label{tab:classification_results}
\footnotesize
\begin{tabular}{|c| c| c| c| c| c|}
\hline
$\theta/\sigma_\Delta$ &
$\rho_E$ &
Event dens. &
Reduction &
Acc. &
Macro-F1\\
 & & (\%) & ($\times$) & (\%) & (\%)\\
\hline
1 & 1.000 & 21.77 & 1.00 & 94.82 & 86.94\\
2 & 0.500 &  3.45 & 6.31 & 93.45 & 79.25\\
5 & 0.200 &  3.15 & 6.91 & 93.37 & 78.82\\
8 & 0.125 &  2.87 & 7.59 & 94.14 & 82.91\\
\hline
\end{tabular}
\end{table}

Table~\ref{tab:classification_results}  evaluates the information-sufficiency boundary $\theta_{\mathrm{IS}}$, the peak-information threshold
$\theta_{I_{\max}}$, the trial-and-error threshold $\theta_{trail}$, and the proposed information-retention threshold $\theta^{0.75}$. At $\theta_{\mathrm{IS}}=\sigma_\Delta$, the classifier achieves 94.82\% accuracy with an event density of 21.77\%. Increasing the threshold to $\theta_{I_{max}} =2\sigma_\Delta$, reduces the event density to 3.45\%, corresponding to a reduction by a factor of 6.31, while the classification accuracy remains 93.45\%. At the operating point selected using~\ref{eq:opt}, $\theta^{0.75}=8\sigma_\Delta$, the event density decreases to 2.87\%, corresponding to a reduction by a factor of 7.59 relative to the information-sufficiency boundary, while the overall classification accuracy is 94.14\%. Thus, for this application, the information-based design identifies a substantially sparser representation without a corresponding reduction in aggregate classification accuracy.

Because aggregate accuracy can obscure class-dependent performance, Table~\ref{tab:classification_results} also reports macro-F1, which assigns equal weight to each class. In the current evaluation, macro-F1 ranges from 78.82\% to 86.94\% across the tested thresholds. At the selected \(8\sigma_\Delta\) operating point, the classifier achieves 82.91\% macro-F1 together with 94.14\% overall accuracy and the lowest event density of 2.87\%. The recovery in both accuracy and macro-F1 at $\theta^{0.75}$ relative to the intermediate thresholds suggests that additional sparsification suppresses event activity that is not beneficial to classification while retaining sufficient class-discriminative information. Because the classifier operates on the full temporal event sequence whereas MI is estimated from a compact activity representation, classification performance need not vary monotonically with the estimated MI.

Importantly, the threshold is selected from the statistical characteristics of the available data and a prescribed information requirement, rather than by repeatedly training and testing the classifier. The normal-condition variability $\sigma_\Delta$ establishes the data-dependent threshold scale, while $\eta$ allows the designer to specify how much of the available class information should be retained when trading information for event sparsity. The methodology therefore provides a systematic way to narrow the threshold design space before classifier training, rather than selecting the encoder threshold through trial-and-error evaluation of task-level accuracy metrics.

\section{Conclusion}
\label{sec:conclusion}
This paper developed an information-theoretic framework for threshold-based event encoding. A formal isomorphism with discretized learning systems identifies a dimensionless variability-to-threshold ratio and an information-sufficiency boundary, while an MI-based criterion allows the encoding threshold to be selected according to an application-defined information-retention requirement $\eta$. For the power-system application, setting $\eta=0.75$ resulted in $\theta^{\eta}=8\sigma_\Delta$. At this operating point, event density was reduced by a factor of approximately 7.6 relative to the information-sufficiency boundary $\theta_{\mathrm{IS}}$, while maintaining comparable classification performance. Thus, rather than prescribing a fixed threshold, the framework provides a data-dependent procedure in which the normal-condition variability establishes the threshold scale and $\eta$ controls the required information retention before classifier training. Future work will investigate application-dependent selection of $\eta$ and extend the analysis to other event-driven learning architectures.


\balance 
\bibliographystyle{IEEEtran}
\bibliography{Ref}

@inproceedings{shukla2026finite,
  author    = {Shukla, Pratishtha and Nutaro, James},
  title     = {Finite-Sample Limits of Entropy-Based Structure
               Identification in Discretized Nonlinear Systems},
  booktitle = {Proceedings of the IEEE Conference on Decision
               and Control},
  year      = {2026},
  note      = {to appear},
  url       = {https://arxiv.org/abs/2609.03074},
}

@inproceedings{alam2024convolutional,
  title={Convolutional Variational Autoencoder-based Unsupervised Learning for Power Systems Faults},
  author={Alam, Maksudul and Yoginath, Srikanth B and Snyder, Isabelle and Winstead, Chris and Stenvig, Nils},
  booktitle={IECON 2024-50th Annual Conference of the IEEE Industrial Electronics Society},
  pages={1--6},
  year={2024},
  organization={IEEE}
}

@article{nilsson2023integration,
  title={Integration of neuromorphic AI in event-driven distributed digitized systems: Concepts and research directions},
  author={Nilsson, Mattias and Schel{\'e}n, Olov and Lindgren, Anders and Bodin, Ulf and Paniagua, Cristina and Delsing, Jerker and Sandin, Fredrik},
  journal={Frontiers in Neuroscience},
  volume={17},
  pages={1074439},
  year={2023},
  publisher={Frontiers Media SA}
}

@article{yao2024spike,
  title={Spike-based dynamic computing with asynchronous sensing-computing neuromorphic chip},
  author={Yao, Man and Richter, Ole and Zhao, Guangshe and Qiao, Ning and Xing, Yannan and Wang, Dingheng and Hu, Tianxiang and Fang, Wei and Demirci, Tugba and De Marchi, Michele and others},
  journal={Nature communications},
  volume={15},
  number={1},
  pages={4464},
  year={2024},
  publisher={Nature Publishing Group UK London}
}

@article{m2025comparative,
  title={A comparative review of deep and spiking neural networks for edge AI neuromorphic circuits},
  author={M Ferreira, Pietro and Wang, Siqi and Gao, Yueyuan and Benlarbi-Delai, Aziz},
  journal={Frontiers in Neuroscience},
  volume={19},
  pages={1676570},
  year={2025},
  publisher={Frontiers}
}

@article{cimarelli2025hardware,
  title={Hardware, algorithms, and applications of the neuromorphic vision sensor: A review},
  author={Cimarelli, Claudio and Millan-Romera, Jose Andres and Voos, Holger and Sanchez-Lopez, Jose Luis},
  journal={Sensors (Basel, Switzerland)},
  volume={25},
  number={19},
  pages={6208},
  year={2025}
}

@article{gallego2020event,
  title={Event-based vision: A survey},
  author={Gallego, Guillermo and Delbr{\"u}ck, Tobi and Orchard, Garrick and Bartolozzi, Chiara and Taba, Brian and Censi, Andrea and Leutenegger, Stefan and Davison, Andrew J and Conradt, J{\"o}rg and Daniilidis, Kostas and others},
  journal={IEEE transactions on pattern analysis and machine intelligence},
  volume={44},
  number={1},
  pages={154--180},
  year={2020},
  publisher={IEEE}
}

@article{khodayar2023deep,
  title={Deep neural networks in power systems: A review},
  author={Khodayar, Mahdi and Regan, Jacob},
  journal={Energies},
  volume={16},
  number={12},
  pages={4773},
  year={2023},
  publisher={MDPI}
}

@article{tian2023neurocare,
  title={NeuroCARE: A generic neuromorphic edge computing framework for healthcare applications},
  author={Tian, Fengshi and Yang, Jie and Zhao, Shiqi and Sawan, Mohamad},
  journal={Frontiers in Neuroscience},
  volume={17},
  pages={1093865},
  year={2023},
  publisher={Frontiers Media SA}
}

@article{pfeiffer2018deep,
  title     = {Deep Learning With Spiking Neurons: Opportunities and Challenges},
  author    = {Pfeiffer, Michael and Pfeil, Thomas},
  journal   = {Frontiers in Neuroscience},
  volume    = {12},
  pages     = {774},
  year      = {2018},
  publisher = {Frontiers Media SA}
}

@article{eshraghian2023snntorch,
  author  = {Eshraghian, Jason K. and Ward, Max and Neftci, Emre O. and Wang, Xinxin and Lenz, Gregor and Dwivedi, Girish and Bennamoun, Mohammed and Jeong, Doo Seok and Lu, Wei D.},
  title   = {Training Spiking Neural Networks Using Lessons From Deep Learning},
  journal = {Proceedings of the IEEE},
  volume  = {111},
  number  = {9},
  pages   = {1016--1054},
  year    = {2023}
}

\end{document}